%% file: ar2.tex
\documentclass[journal,twocolumn]{IEEEtran}
\usepackage{amsfonts, amsmath,amssymb, graphicx, algorithm,algpseudocode,color,cite,epstopdf,subfigure,cite,bm,multirow,booktabs}

\include{def}

\allowdisplaybreaks

\graphicspath{figure}

\begin{document}
\title{Amplitude Retrieval for Channel Estimation of MIMO Systems with One-Bit ADCs}
\author{Cheng Qian, \emph{Member, IEEE}, Xiao Fu, \emph{Member, IEEE}, and Nicholas D. Sidiropoulos, \emph{Fellow, IEEE}
\thanks{This work was supported in part by the National Science Foundation under project NSF ECCS-1807660 and NSF ECCS-1608961.\par
	C. Qian and N. D. Sidiropoulos are with the Department of Electrical and Computer Engineering, University of Virginia, Charlottesville, VA 22904 USA (e-mail: alextoqc@gmail.com, nikos@virginia.edu).\par
	X. Fu is with School of Electrical Engineering and Computer Science, Oregon State University, Corvallis, OR 97331 (xiao.fu@oregonstate.edu).
}	
}

\maketitle

\begin{abstract}
	 This letter revisits the channel estimation problem for MIMO systems with one-bit analog-to-digital converters (ADCs) through a novel algorithm--\emph{Amplitude Retrieval (AR)}. Unlike the state-of-the-art methods such as those based on one-bit compressive sensing, AR takes a different approach. It accounts for the lost amplitudes of the one-bit quantized measurements, and performs channel estimation and amplitude completion jointly. This way, the direction information of the propagation paths can be estimated via accurate direction finding algorithms in array processing, e.g., maximum likelihood. The upsot is that AR is able to handle off-grid angles and provide more accurate channel estimates. Simulation results are included to showcase the advantages of AR.
\end{abstract}

\vspace{-0.8em}
\section{Introduction}

An effective way to mitigate hardware complexity, cost, and power consumption in commercial large-scale multiple-input multiple-output (MIMO) systems is to use one-bit analog-to-digital converters (ADCs) at both transmitter and receiver. However, with such configuration, each received data sample gives rise to a single bit, i.e., only the sign of that sample is recorded. This challenges many standard channel estimation techniques such as least squares and minimum mean-square error, which were originally designed for finely quantized data. Therefore, developing novel channel estimation algorithms for MIMO systems with one-bit ADCs is of practical and methodological interest.

In the past, numerous successes have been reported on channel estimation from sign measurements, such as \cite{mo2014channel,choi2016near,wang2014multiuser,jacobsson2015one}. A common assumption behind these algorithms is that the multipath channel consists of a few distinct paths.  Most of these algorithms are compressive sensing (CS) based methods that exploit the sparsity of the channel in the angular domain for channel estimation. Following this theme, the authors of \cite{mo2014channel} proposed an expectation-maximum (EM) channel estimator. In \cite{choi2016near}, a near maximum likelihood method was devised and shown to attain better performance than the EM algorithm. These CS based methods perform well under favorable conditions -- perfect sparsity, known number of paths, and high signal-to-noise ratio (SNR) -- e.g., see \cite{mo2014channel}. However, such clean and well-determined scenarios rarely arise in practice. One major concern arises from the over-complete dictionaries. Under the double-directional channel model, the actual dictionary is a Kronecker product of two or more ``fat'' dictionary matrices, which results in a very hard linear inverse problem. Moreover, to handle random direction-of-arrival (DOA) and direction-of-departure (DOD), fine grids should be used to generate each dictionaries. Therefore, the final dictionary has strong correlation between the adjacent columns, which further aggrevates the parameter estimation problem. 

The work in \cite{alevizos2018limited,zhou2017sparse} studied a similar channel estimation problem but from a different perspective, where downlink channel estimation is performed at the base station using bits fed back from mobile users, conveying just the sign of their received signal. These methods can be directly applied to solve the channel estimation problem for MIMO systems with one-bit ADCs. In \cite{alevizos2018limited}, the authors exploited the sparsity in the channel matrix and developed an efficient algorithm with closed-form expression for channel estimation. The algorithm works with a nonuniform dictionary which requires some prior knowledge of the DOA/DOD information of the paths that may not be feasible in practice, especially in fast moving scenarios where DOA/DOD could change rapidly. The method in \cite{zhou2017sparse} is based on the so-called binary iterative hard thresholding
(BIHT) technique \cite{jacques2013robust}. Both \cite{alevizos2018limited} and \cite{zhou2017sparse} suffer from the same drawbacks as the aforementioned one-bit CS based channel estimators.

Can we avoid the inherent limitations of overcomplete, highly correlated dictionaries and one-bit CS, enabling high-resolution estimates of the path angles, even if off-grid?  Our goal in this letter is to obtain an affirmative answer to this question.  
We propose a novel algorithm named {\em amplitude retrieval} (AR) for downlink channel estimation from only sign measurements. Unlike pre-existing methods, AR takes the lost amplitudes of the sign measurements into account and alternately optimizes between the amplitudes and channel parameters. The advantages in algorithm design of doing this are two-fold: 1) It avoids the construction of large-dimensional dictionary, and it does not suffer performance degradation caused by the coherency in the dictionary;  and 2) it enables to handle off-grid angles via more sophisticated algorithms such as maximum likelihood (ML) for multidimensional harmonic retrieval. In this way, we can simultaneously improve the estimation accuracy of channel parameters, i.e., DOA, DOD and path gain. Simulation results are included to show the effectiveness of AR.


\section{Signal Model}
Consider the downlink of a narrowband (or a single tone in a wideband OFDM) FDD MIMO system equipped with one-bit ADCs on the receiver end, where the recorded signal is
\begin{align}\label{Y}
\Y = \sign\(\Re(\H\S + \N)\) + j\sign\(\Im(\H\S + \N)\)
\end{align}
where $\sign(\cdot)$, $\Re(\cdot)$ and $\Im(\cdot)$ take the sign, real and imaginary parts of their arguments, respectively, $\S\in\bC^{M_t\times N}$ denotes the training signal, $\N\in\bC^{M_r\times N}$ is white Gaussian noise, and $\H\in\bC^{M_r\times M_t}$ is the channel matrix. Here, $N$, $M_r$ and $M_t$ denote the number of training samples, receive antennas and transmit antennas, respectively. 

We assume a specular multipath channel comprising a few dominant paths, each characterized by DOA, DOD and channel gain [4]–[7]. This channel model is appropriate under certain conditions. For example, when the BS antenna array is mounted on top of a tall building or cellular tower, such that the number of local scatterers is limited; or when the carrier frequency is lifted to the millimeter wave range, where due to the severe path loss only a few specular reflections arrive at the other end of the link \cite{heath2016overview,xie2016overview}.
Based on this assumption, the channel matrix can be expressed as
\begin{align}\label{H}
\H = \A_r(\bth)\diag(\bbe)\A_t^H(\bph)
\end{align}
where $(\cdot)^H$ is the conjugate transpose, $\diag(\cdot)$ denotes a diagonal matrix, $\A_r(\bth) = [\a_{r}(\theta_1)\ \cdots\ \a_{r}(\theta_K)]$, $\A_t(\bph) = [\a_{t}(\phi_1)\ \cdots\ \a_{t}(\phi_K)]$ and $\bbe = [\beta_1\ \cdots\ \beta_K]^T$. Here, $(\cdot)^T$ is the transpose, $\{\theta_k\}$, $\{\phi_k\}$ and $\{\beta_k\}$ are DOA, DOD and path-loss, respectively, $\a_r(\theta_k) = \[ e^{j2\pi f_0\tau_1(\theta_k)}\ \cdots \ e^{j2\pi f_0\tau_{M_r}(\theta_k)} \]^T$ and $\a_t(\phi) = \[ e^{j2\pi f_0\tau_1(\phi)}\ \cdots \ e^{j2\pi f_0\tau_{M_t}(\phi)} \]^T$ are the $k$th steering vector at the receiver and transmitter, respectively, in which $\tau_m(\theta_k)$ is the time needed by the signal that bounces off the $k$th scatterer to reach the $m$th receive antenna, and $\tau_m(\phi_k)$ is the time that the signal emitted from the $m$th transmit antenna takes to arrive at the $k$th scatterer.

\subsection{Challenges} 
In FDD systems, to control the downlink training overhead, the training signal $\S$ could be a tall matrix, meaning that $\S$ cannot be canceled out using matched filtering technique. This poses an extremely hard problem for channel parameter estimation, especially for DOD estimation. To handle this issue, one seeks to exploit the sparsity in the angle domain, i.e., constructing two dictionaries for DOA and DOD, respectively, and then transforming the channel estimation problem to an one-bit compressive sensing problem. 
Several algorithms have been proposed based on this idea \cite{mo2014channel,choi2016near,wang2014multiuser,jacobsson2015one,alevizos2018limited,zhou2017sparse}. However, one major concern in these approaches is that they might be impotent when dealing with off-grid angles due to the limited resolution of their angular dictionary. To ensure high resolution, fine grids can be used to construct the dictionary, but this  inevitably causes coherency in adjacent columns of the dictionary and ultimately leads to unsatisfactory performance.
Furthermore, due to the lack of amplitude information about $\H\S$, there is no uniqueness guarantee of the channel estimate, even at high SNR, and any scaled version of $\H$ which does not change the signs is admissible as well. A common way to fix this issue is to enforce a norm constraint, i.e., $\|\H\|_F^2=R$, where $R$ is an expected value of the channel norm. Unfortunately, such a constraint is nonconvex w.r.t. $\H$, which makes optimization design more challenging.

\section{Amplitude Retrieval}
In this section, we derive an AR algorithm to alleviate the aforementioned challenges. Before proceeding with the detailed  derivations, let us first bring forth our motivation. We note that the data matrix $\Y$ only preserves the signs of the real and imaginary parts of the baseband radio signals, while the corresponding amplitudes are lost. 
This can be viewed the `dual' of the classical phase retrieval problem \cite{candes2015phase}, in which the phase of the received data is missing and the signal of interest can be efficiently estimated by completing the phase \cite{wang2017solving,qian2017inexact}.
Assuming that the amplitude of $\Y$, denoted by $\bGa$, is available, then we have, in the noiseless case
\begin{equation}
\Y\circledcirc\bGa = \H\S
\end{equation} 
where $\circledcirc$ is defined as $a\circledcirc b = \Re(a)\circledast\Re(b)+j\Im(a)\circledast\Im(b)$ with $\circledast$ being the element-wise product. Estimating $\H$ from $\Y\circledcirc\bGa$ is a relatively easy task compared to the channel estimation from $\Y$. 
Based on this motivation, we propose a novel channel estimation algorithm that we call \emph{Amplitude Retrieval (AR)}, which aims to estimate the lost amplitude and channel parameters jointly.

AR deals with the following optimization problem
\begin{align}\label{prob3}
\min_{\H,\bth,\bbe,\bph,\bGa}\,& \left\|\Y\circledcirc\bGa - \H\S\right\|_F^2 + \lambda\left\| \H - \A_r(\bth)\diag(\bbe)\A_t^H(\bph) \right\|_F^2 \notag\\
\text{s. t.}\quad\, & \Re(\bGa)\geq 0,\,\Im(\bGa)\geq 0,~\|\H\|_F^2=R
\end{align}
where $\|\cdot\|_F$ is the Frobenius norm, the first two constraints ensure non-negativity of the amplitude, and the third constraint is to prevent the scaling ambiguity. It should be highlighted that Problem \eqref{prob3} does not rely on any angular dictionary and it enables to handle off-grid angles using an ML method. This is a major advantage over many existing compressive sensing based approaches.

We adopt alternating optimization to solve \eqref{prob3}, i.e., we optimize over $\{\bGa,\{\bth,\bbe,\bph\}\}$ and $\H$ by fixing one for another. Assume that after $r$ iterations, we have a channel estimate $\H^{(r)}$ available. Then at the next iteration, the subproblem w.r.t. $\bGa$ is given by
\begin{align}\label{prob:Gamma}
\min_{\bGa}\;& \big\|\Y\circledcirc\bGa - \H^{(r)}\S\big\|_F^2 \notag\\
\text{s. t.}\; & \Re(\bGa)\geq 0,\Im(\bGa)\geq 0
\end{align}
which is convex and can be optimally solved in closed form:
\begin{align}\label{update:Gamma}
\bGa^{(r+1)} =&\; \max\( \Re\(\Y\)\circledast\Re(\H^{(r)}\S), 0 \) + \notag\\
& ~j\max\( \Im\(\Y\)\circledast\Im(\H^{(r)}\S), 0 \)
\end{align}
where $\max(a,b)=a$ if $a\geq b$; otherwise, $\max(a,b)=b$.

The subproblem for $\{\bth,\bbe,\bph\}$ is
\begin{align}\label{prob:theta}
\min_{\bth,\bbe,\bph}\; \left\| \H^{(r)} - \A_r(\bth)\diag(\bbe)\A_t^H(\bph) \right\|_F^2.
\end{align}
Note that when the transmit and receive antennas are uniform linear arrays (ULAs) or uniform rectangular arrays (URAs), \eqref{prob:theta} becomes a multi-dimensional harmonic retrieval (HR) problem. Existing algorithms such as\cite{qian2018robust} and ML \cite{bresler1986exact} are applicable to handle \eqref{prob:theta}. However, when non-uniform arrays are used, HR methods are no longer applicable. Instead, we can seek an ML method to optimize \eqref{prob:theta}.
Towards this goal, let us rewrite the objective function in \eqref{prob:theta} as
$$
\left\| \H^{(r)} - \A_r(\bth)\diag(\bbe)\A_t^H(\bph) \right\|_F^2 = \left\| \h^{(r)} - \A(\bth,\bph)\bbe \right\|_2^2
$$
where $\|\cdot\|_2$ is the $\ell_2$-norm, $\h^{(r)} = \text{vec}\big(\H^{(r)}\big)$ and $\A(\bth,\bph) = \A_t^*(\bph)\odot\A_r(\bth)$ with $(\cdot)^*$, $\odot$ and $\text{vec}(\cdot)$ denoting the complex conjugate, Khatri-Rao product and vectorization operator, respectively.

It follows from \cite{stoica1988music} that substituting the least squares estimate of $\bbe$, i.e., $\hat\bbe = \A^\dagger\h^{(r)}$ 
into \eqref{prob:theta}, the ML cost for $\{\bth,\bph\}$ is obtained as 
$f(\bet) = \big\| \P_\A^\perp\h^{(r)} \big\|_2^2$ 
where $(\cdot)^\dagger$ is the pseudo-inverse, $\bet = \big[~\bth^T\; \bph^T~\big]^T$ and $\P_\A^\perp = \I - \A\A^\dagger$ denotes the orthogonal projection with $\I$ being an identity matrix.
We employ a few gradient iterations to update $\bet$, i.e.,
\begin{align}\label{gradient:eta}
\bet^{(\ell+1)} = \bet^{(\ell)} - \mu^{(\ell)}\pmb{\nabla} J(\bet^{(\ell)})
\end{align}
where $\mu^{(\ell)}$ is the step-size and
\begin{align*}
\pmb{\nabla}f(\bet) = -2\,\Re\left(
\begin{bmatrix}
\(\A^\dagger\h^{(r)}\) \circledast \(\big(\h^{(r)}\big)^H\P_\A^\perp\D(\bth)\)^T\\
\(\A^\dagger\h^{(r)}\) \circledast \(\big(\h^{(r)}\big)^H\P_\A^\perp\D(\bph)\)^T
\end{bmatrix}
\right).
\end{align*}
When $\bet$ is updated, $\bbe$ is subsequently estimated as
\begin{align}\label{beta}
\bbe^{(r+1)} = \(\A\big(\bth^{(r+1)},\bph^{(r+1)}\big)\)^\dagger\h^{(r)}.
\end{align}

With $\bGa^{(r+1)}$ and $\big\{ \bth^{(r+1)},\bbe^{(r+1)},$ $\bph^{(r+1)} \big\}$ available, the subproblem with respect to $\H$ becomes
\begin{align}\label{prob:H}
\min_{\H}\;& \left\|\Y\circledcirc\bGa^{(r+1)} - \H\S\right\|_F^2 + \lambda\left\| \H - \tilde{\H}^{(r+1)} \right\|_F^2 \notag\\
\text{s. t.}\; &\|\H\|_F^2=R
\end{align}
where
\begin{align}\label{update:C}
\tilde{\H}^{(r+1)} = \A_r(\bth^{(r+1)})\diag(\bbe^{(r+1)})\A_t^H(\bph^{(r+1)}).
\end{align}
The above is a quadratically constrained quadratic programming (QCQP) problem with one constraint. It is well known that for such a QCQP problem, even though the constraint is non-convex, it can always be solved to optimality, by strong duality \cite{huang2016consensus,tao1998dc}.
In the following derivations, the superscript $(r+1)$ will be temporally removed for notational simplicity.

First, let us write down the Lagrangian of \eqref{prob:H} as \cite{tao1998dc}
\begin{align}\label{prob:H2}
\!\!\mathcal{L} =  \|\Y\circledcirc\bGa- \H\S\|_F^2 + \lambda\| \H - \tilde{\H} \|_F^2  + \rho (\|\H\|_F^2-R)
\end{align}
where $\rho$ is the dual variable. The necessary condition for optimality is that $\nabla \mathcal{L}=0$, i.e.,
\begin{align}
\nabla \mathcal{L} = 
\H\big(\S\S^H + \tilde{\rho}\I\big) - \bLa = 0
\end{align}
which results in
\begin{align}\label{H2}
\H &= \bLa\(\S\S^H + \tilde{\rho}\I\)^{-1}
\end{align}
where $(\cdot)^{-1}$ is matrix inverse, $\tilde{\rho} = \rho + \lambda$, $\bLa = (\Y\circledast\bGa)\S^H+\lambda\tilde{\H}$. Now the problem is to find $\tilde{\rho}$ such that \eqref{H2} is feasible.

Substituting \eqref{H2} into the norm constraint in \eqref{prob:H} yields
\begin{align}\label{prob:rho}
\tr\( \bLa\(\S\S^H + \tilde{\rho}\I\)^{-2}\bLa^H \)=R
\end{align}
which equals to
\begin{align}\label{prob:varsigma}
\sum_{i=1}^{M_t} \frac{\|\c_1\|_2^2}{(\tilde{\rho} + \varsigma_i)^2} = R
\end{align}
where $\tr(\cdot)$ is the trace operator, $\c_i$ is the $i$th column of $\bLa\U^H$ with $\U$ being the eigenvector matrix of $\S\S^H$, and $\varsigma_i$ is the corresponding eigenvalue. Since $\{\c_i,\varsigma_i\}$ are all known, $\rho$ can be numerically found by solving \eqref{prob:varsigma}.

In the following, we briefly show that there exists a real  $\tilde{\rho}$ such that \eqref{prob:rho} holds true. Define $f(\tilde{\rho}) = \sum_{i=1}^{M_t} \frac{\|\c_1\|_2^2}{(\tilde{\rho} + \varsigma_i)^2} - R$. 
Then 
\begin{align}
f'(\tilde{\rho}) = -2\sum_{i=1}^{M_t}\frac{\|\c_i\|_2^2}{(\tilde{\rho}+\varsigma_i)^{3}}.
\end{align}

It follows from \cite{huang2016consensus} that for the dual of QCQP with one constraint, we have $\S\S^H + \tilde{\rho}\I\succ0$. Therefore, $f'(\tilde{\rho})< 0$, meaning that $f(\tilde{\rho})$ is monotonically decreasing and such feasible $\tilde{\rho}$ is unique. Moreover, $\tilde{\rho}+\varsigma_i>0,\forall i$ implies that $\varsigma_{\min}<\tilde{\rho}<+\infty$, 
therefore \eqref{prob:rho} always has a root in $(\varsigma_{\min},+\infty)$, where $\varsigma_{\min}$ is the smallest eigenvalue of $\S\S^H$. We may employ either bisection or Newton's method to estimate $\tilde{\rho}$. Once $\tilde{\rho}$ is found, plugging it back to \eqref{H2}, we obtain the optimal solution for \eqref{prob:H}. 

\subsubsection*{Special Case I}
When $\S$ semi-unitary, i.e., $\S^H\S=\I$, using the matrix inversion lemma, we can rewrite 
\eqref{prob:rho} as
\begin{align}\label{15}
\!\!\!R\tilde{\rho}^4 + 2R\tilde{\rho}^3 + (R - t_1)\tilde{\rho}^2 + 2(t_2 - t_1)\tilde{\rho} + t_2 - t_1 = 0
\end{align}
where $t_1 = \tr(\bLa\bLa^H)$ and $t_2 = \tr(\bLa\S\S^H\bLa^H)$.
In this case, $\varsigma_{\min}=0$ and $\tilde{\rho}>0$, which implies that \eqref{15} has a unique positive root which is indeed the right $\tilde{\rho}$. Furthermore, we can also avoid the inverse in \eqref{H2} to reduce the complexity in the update of $\H$, i.e., $\H = \bLa\( \I/\tilde{\rho} - \S\S^H/(\tilde{\rho}(1+\tilde{\rho})) \)$.

\subsubsection*{Special Case II}
When $\S$ unitary, i.e., $\S^H\S=\S\S^H=\I$. Eq. \eqref{prob:rho} becomes $ (1 + \tilde{\rho})^2 = \tr(\bLa\bLa^H)/R$. 
Then the optimal solution is
\begin{align}\label{16}
\tilde{\rho} = \sqrt{\tr(\bLa\bLa^H)/R} - 1.
\end{align}

Algorithm \ref{Alg:AR} summarizes the detailed steps of AR.
\begin{algorithm}[h!]
	\caption{AR for channel estimation}\label{Alg:AR}
	\begin{algorithmic}[1]
		\State Set $r = 1$
		\While{stopping criterion has not been reached}
		\State Compute $\bGa^{(r)}$ via \eqref{update:Gamma}
		\State Update $\big\{\bth^{(r)},\bbe^{(r)},\bph^{(r)}\big\}$ using \eqref{gradient:eta}
		\State Refine $\H^{(r)}$ through \eqref{H2}, where $\tilde{\rho}$ is calculated by solving \eqref{prob:rho}, or through \eqref{15} when $\S$ is semi-unitary and \eqref{16} when $\S$ is unitary
		\State $r=r+1$
		\EndWhile
	\end{algorithmic}
\end{algorithm}

\begin{remark}
	We emphasize that although our method is presented in a downlink setting, its usage is not limited to that. Actually, AR can be easily extended to solve the uplink channel estimation problem under the TDD protocol without changing anything, as we will see in Section IV. 
\end{remark}

\section{Performance Comparison}
In this section, we present numerical results to evaluate the performance of AR by comparing it with the state-of-the-art algorithms including a soft-thresholding (ST) based technique \cite{alevizos2018limited}, joint BIHT (J-BIHT) \cite{zhou2017sparse} and the EM algorithm \cite{mo2014channel}. We report their mean square error (MSE) performance which is defined as 
$\text{MSE} = 1/100\sum_{i=1}^{200} \big\| \hat{\H_i}/\|\hat{\H_i}\|_F - \H_i/\|\H_i\|_F \big\|_F^2$. Note that for ST, EM and J-BIHT, their dictionary is with 7 bits, i.e., we split $[0, \pi]$ using 128 grids points. Therefore, the minimum resolution ability using this dictionary is $\pi/128$. In the following simulations, by taking this resolution into account,  we consider relatively widely-spaced DOAs/DODs, i.e., any two adjacent DOAs or DODs are separated at least $\pi/16$. 

\subsection{Downlink Channel Estimation in FDD}
\begin{figure}
	\centering
	\includegraphics[width=0.9\linewidth]{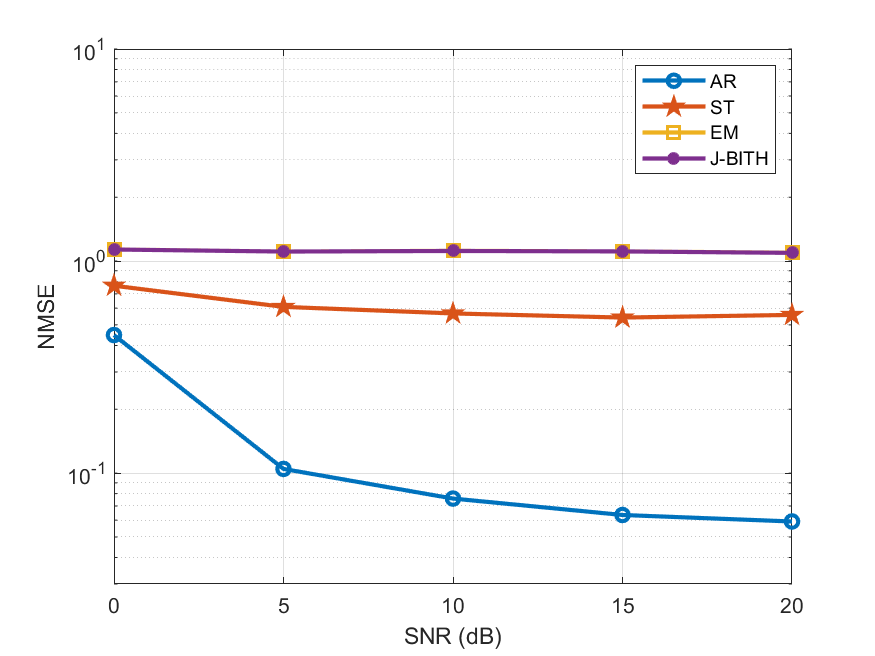}
	\caption{NMSE vs SNR (downlink case)}
	\label{fig:ulamsesnr}
\end{figure}

\begin{figure}
	\centering
	\includegraphics[width=0.9\linewidth]{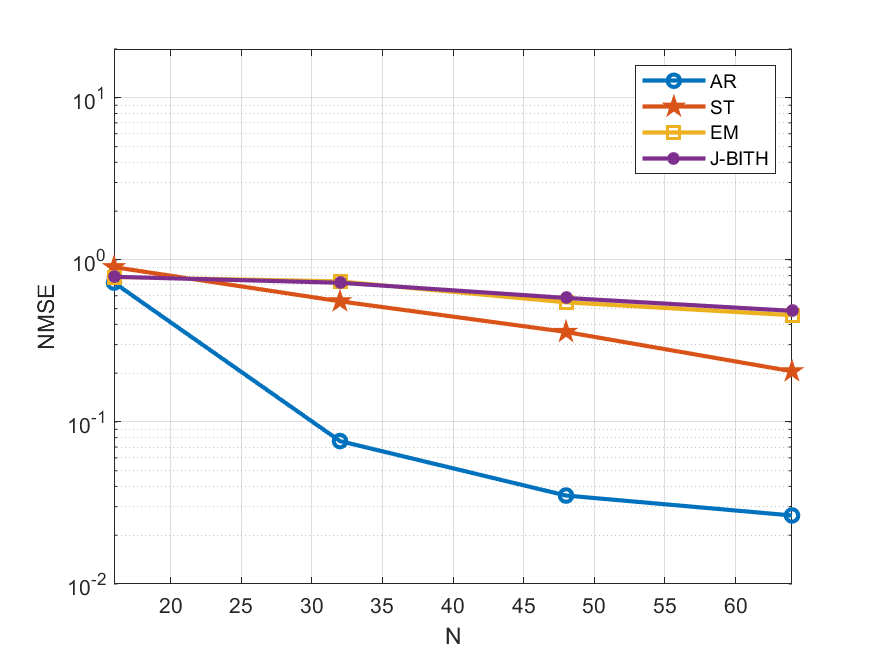}
	\caption{NMSE vs $N$ (downlink case)}
	\label{fig:ulamsen}
\end{figure}

We consider a downlink case, where the transmit and receive antenna arrays are ULAs with number of elements being $M_r=4$ and $M_t=64$, respectively. 
The number of training samples is $N=32$. We fix the number of paths as $K=5$ and generate angles from the range $[0, \pi]$. Fig. \ref{fig:ulamsesnr} shows the NMSE performance as a function of SNR. We can see that AR achieves substantially better performance than the other three methods. EM and J-BITH failed to perform well. The main reason is that the two algorithms intrinsically assumed on-grid DOAs/DODs and an overdetermined setting, i.e., the number of paths is smaller than $M_r$. However, in this example, the number of paths is greater than $M_r$, which violates the overdetermined assumption. Similar results can also be found in Fig. \ref{fig:ulamsen}, where $\mathrm{SNR}=10$ dB and the way of generating channel parameters is the same as in Fig. \ref{fig:ulamsesnr}. As $N$ increases, the performance of the competitors increases accordingly. AR still performs the best and ST is the second best.

\subsection{Uplink Channel Estimation in TDD}
\begin{figure}
	\centering
	\includegraphics[width=0.9\linewidth]{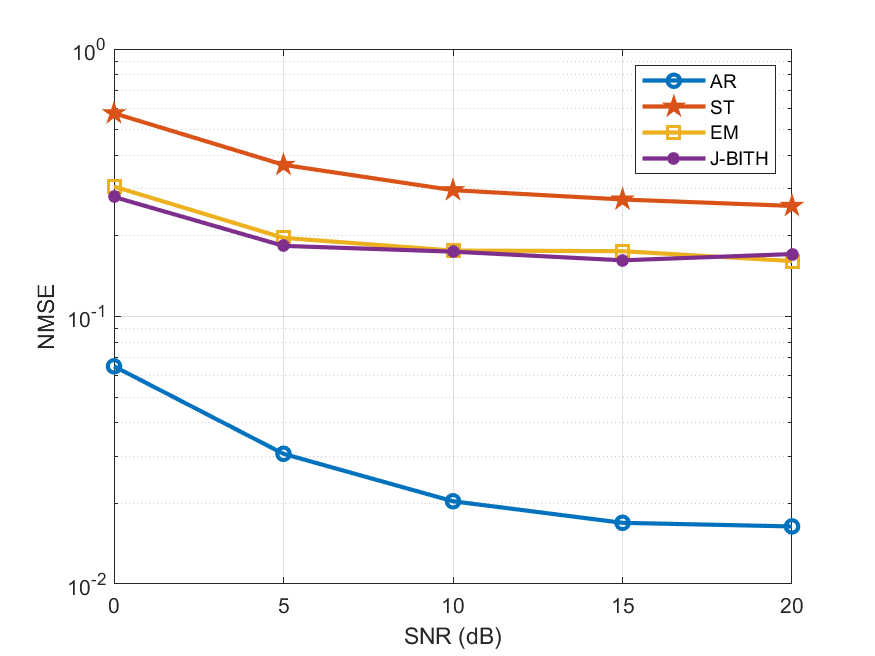}
	\caption{NMSE versus SNR (uplink case)}
	\label{fig:tddulamsesnr}
\end{figure}

As we have mentioned in Section III, AR can be used for uplink channel estimation in TDD systems. To verify this, we define a simple uplink scenario, where the BS has a ULA with 64 antennas receiving signals from $K=16$ users. Here, we assume that each user is equipped with a single antenna and communicates with the BS through one dominant propagation path. Moreover, orthogonal training signals are used, i.e., $N=K$ and $\S\S^H=\I$. The NMSE performance versus SNR is plotted in Fig. \ref{fig:tddulamsesnr}. AR is the best in terms of NMSE performance. However, unlike the downlink case, the EM and J-BIHT algorithms showed much better performance. This is because the number of users is much smaller than the number of the antennas at the BS, which is an overdetermined setting and meets the assumption in EM and J-BIHT.

\section{Conclusion}
We have devised an AR algorithm for MIMO channel estimation from one-bit quantized measurements. AR is able to avoid the shortcomings of many CS based algorithms. It alternately optimizes between the amplitudes of the measurements and channel parameters. With completed amplitudes, the DOAs, DODs and path gains can be estimated using harmonic retrieval or ML  algorithms. Thus, AR can deal with any random propagation path, not necessarily on a grid. Simulations have verified the success of AR relative to the state-of-art in the literature. 


\bibliographystyle{IEEEtran}
\bibliography{IEEEabrv,mybib}

\end{document}

%% file: def.tex
\def\Re{\text{Re}}
\def\Im{\text{Im}}

\def\bbe{\boldsymbol{\beta}}

\def\bet{\boldsymbol{\eta}}
\def\bth{\boldsymbol{\theta}}

\def\bph{\boldsymbol{\phi}}

\def\bGa{\mathbf{\Gamma}}

\def\bLa{\mathbf{\Lambda}}

\def\tr{\text{trace}}
\def\({\left(}
\def\){\right)}
\def\[{\left[\,}
\def\]{\,\right]}

\def\0{\boldsymbol{0}}
\def\1{\boldsymbol{1}}
\def\a{\mathbf{a}}
\def\A{\mathbf{A}}

\def\c{\mathbf{c}}

\def\bC{\mathbb{C}}

\def\D{\mathbf{D}}

\def\h{\mathbf{h}}
\def\H{\mathbf{H}}

\def\I{\mathbf{I}}

\def\N{\mathbf{N}}

\def\P{\mathbf{P}}

\def\S{\mathbf{S}}

\def\U{\mathbf{U}}

\def\Y{\mathbf{Y}}

\def\diag{\mathrm{diag}}
\def\sign{\mathrm{sign}}

\newtheorem{remark}{Remark}